Lysine-Cysteine-Serine-Tryptophan Inserted into the DNA-Binding Domain of Human Mineralocorticoid Receptor Increases Transcriptional Activation by Aldosterone

Yoshinao Katsu[1,2,#]  Jiawen Zhang[2,#], Michael E. Baker[3,4,] *

[1] Faculty of Science

Hokkaido University

Sapporo, Japan

[2] Graduate School of Life Science

Hokkaido University

Sapporo, Japan

[3] Division of Nephrology-Hypertension

Department of Medicine, 0693

University of California, San Diego

9500 Gilman Drive

La Jolla, CA 92093-0693

Center for Academic Research and Training in Anthropogeny (CARTA) [4]

University of California, San Diego

La Jolla, CA 92093

\# Equal contribution

*Correspondence to

Y. Katsu; E-mail: ykatsu@sci.hokudai.ac.jp

M. E. Baker; E-mail: mbaker@health.ucsd.edu

**Abstract**

Due to alternative splicing in an ancestral DNA-binding domain (DBD) of the mineralocorticoid receptor (MR), humans contain two almost identical MR transcripts with either 984 amino acids (MR-984) or 988 amino acids (MR-988), in which their DBDs differ by only four amino acids, Lys,Cys,Ser,Trp (KCSW). Human MRs also contain mutations at two sites, codons 180 and 241, in the amino terminal domain (NTD). Together, there are five distinct full-length human MR genes in GenBank. Human MR-984, which was cloned in 1987, has been extensively studied. Human MR-988, cloned in 1995, contains KCSW in its DBD. Neither this human MR-988 nor the other human MR-988 genes have been studied for their response to aldosterone and other corticosteroids. Here, we report that transcriptional activation of human MR-988 by aldosterone is increased by about 50% compared to activation of human MR-984 in HEK293 cells transfected with the TAT3 promoter, while the half-maximal response (EC50) is similar for aldosterone activation of MR-984 and MR-988. Transcriptional activation of human MR also depends on the amino acids at codons 180 and 241. Interestingly, in HEK293 cells transfected with the MMTV promoter, transcriptional activation by aldosterone of human MR-988 is similar to activation of human MR-984, indicating that the promoter has a role in the regulation of the response of human MR-988 to aldosterone. The physiological responses to aldosterone and other corticosteroids in humans with MR genes containing KCSW and with differences at codons 180 and 241 in the NTD warrant investigation.

Running title: Increased Corticosteroid Activation of the Human Mineralocorticoid Receptor containing KCSW in the DBD

**1. Introduction.**

The mineralocorticoid receptor (MR) is a ligand-activated transcription factor that belongs to the nuclear receptor family, which arose in multicellular animals along with other vertebrate steroid receptors: the glucocorticoid receptor (GR), progesterone receptor (PR), androgen receptor (AR) and estrogen receptor (ER) [1–6]. The classical function of the MR in humans and other terrestrial vertebrates is to maintain electrolyte balance by regulating sodium and potassium transport in epithelial cells in the kidney and colon [7–12]. In addition, the MR also has important physiological functions in many other tissues, including brain, heart, skin and lungs [11,13–22].

The human MR sequence reported in 1987 by Arriza et al [1] contains 984 amino acids (MR-984). Inspection of the MR sequence revealed that, like other steroid receptors, the human MR is composed of four modular functional domains: a large amino-terminal domain (NTD) of about 600 amino acids, followed in the center by a DNA-binding domain (DBD) of about 65 amino acids, followed by a small hinge domain of about 60 amino acids that connected to the ligand-binding domain of about 250 amino acids at the C-terminus, where aldosterone, cortisol and other corticosteroids bind to activate transcription [1,2,23–25].

Analysis of the human MR sequence by Arriza et al. [1] revealed that the MR was closely related to the glucocorticoid receptor (GR), which was consistent with evidence that some corticosteroids, such as cortisol, corticosterone and 11-deoxycorticosterone were ligands for both the MR and GR [13,15,21,24,26] and that aldosterone, cortisol, corticosterone and 11-deoxycorticosterone have similar binding affinity for human MR [1,21,26–28]. Activation of the MR by cortisol and corticosterone, two steroids that are ligands for the GR [28–30], is consistent with the evolution of the GR and MR from a common ancestral corticoid receptor (CR) in a cyclostome (jawless fish) that evolved about 550 million years ago at the base of the vertebrate line [24,31–37].

Knowledge about the human MR family expanded in 1995, when Bloem et al. [38] cloned a human MR with 988 amino acids (MR-988). This human MR had four additional amino acids in the DBD due to alternative splicing between exons 3 and 4 of an in-frame insertion of 12 bp encoding Lys, Cys, Ser, Trp (KCSW) [39,40] (Figure 1). As shown in Figure 1, the DBDs in terrestrial vertebrate MRs are highly conserved.

```
Human MR-984:         CLVCGDEASGCHYGVVTCGSCKVFFKRAVEG----QHNYLCAGRNDCIIDKIRRKNCPACRLQKCLQAGM
Human MR-988:         CLVCGDEASGCHYGVVTCGSCKVFFKRAVEGKCSWQHNYLCAGRNDCIIDKIRRKNCPACRLQKCLQAGM
Platypus MR-990:      CLVCGDEASGCHYGVVTCGSCKVFFKRAVEG----QHNYLCAGRNDCIIDKIRRKNCPACRLQKCLQAGM
Platypus MR-994:      CLVCGDEASGCHYGVVTCGSCKVFFKRAVEGKCSWQHNYLCAGRNDCIIDKIRRKNCPACRLQKCLQAGM
Turtle MR-993:        CLVCGDEASGCHYGVVTCGSCKVFFKRAVEG----QHNYLCAGRNDCIIDKIRRKNCPACRLQKCLQAGM
Turtle MR-997:        CLVCGDEASGCHYGVVTCGSCKVFFKRAVEGKCSWQHNYLCAGRNDCIIDKIRRKNCPACRLQKCLQAGM
Xenopus MR-979:       CLVCGDEASGCHYGVVTCGSCKVFFKRAVEG----QHSYLCAGRNDCIIDKIRRKNCPACRLQKCLQAGM
Xenopus MR-983:       CLVCGDEASGCHYGVVTCGSCKVFFKRAVEGKCSRQHSYLCAGRNDCIIDKIRRKNCPACRLQKCLQAGM
Lungfish MR-985:      CLVCGDEASGCHYGVVTCGSCKVFFKRAVEG----QHNYLCAGRNDCIIDKIRRKNCPACRVRKCLQAGM
Zebrafish MR-970:     CLVCGDEASGCHYGVVTCGSCKVFFKRAVEG----QHNYLCAGRNDCIIDKIRRKNCPACRVRKCLQAGM
Elephant shark MR-956:CLVCSDEASGCHYGVLTCGSCKVFFKRAVEG----QQNYLCAGRNDCIIDKIRRKNCPACRLRKCLKAGM
```

**Figure 1. Comparison of the DNA-binding domains on human MR-984, human MR-988, platypus MR-990, platypus MR-994-, turtle MR-993, turtle MR-997, Xenopus MR-979, Xenopus MR-983, lungfish MR-985, zebrafish MR-970, elephant shark MR-956.** The DBD of human MR-988 has an insertion of KCSW that is absent in human MR-984. Otherwise, the rest of the sequences of human MR-984 and human MR-988 are identical. A four amino acid insertion also is present at a corresponding position in the DBD of platypus MR turtle MR, and *Xenopus* MR. Moreover, except for the four amino acid insert in the DBD, the rest of the DBD sequences of human MR, platypus MR and turtle MR are identical. Differences between the DBD sequence in human MR and selected vertebrate MRs are shown in red. Protein accession numbers are XP_054206038 for human MR-984, XP_054206037 for human MR-988, XP_007669969 for platypus MR-990, XP_016083764 for platypus MR-994, XP_044874191 for turtle MR-993, XP_044874189 for turtle MR-997, NP_001084074 for Xenopus MR-979, XP_018098693 for *Xenopus* MR-983, BCV19931 for lungfish MR-985, NP_001093873 for zebrafish MR-970 and XP_007902220 for elephant shark MR-956.

Thus, humans were found to contain two almost identical MR transcripts, in which their DBDs differ by only four amino acids. The number of distinct human MR genes expanded after a BLAST analysis of GenBank with the human MR sequence identified differences at codons 180 and 241 in the NTD of MR-984 and MR-988 in humans [41] (Figure 2).

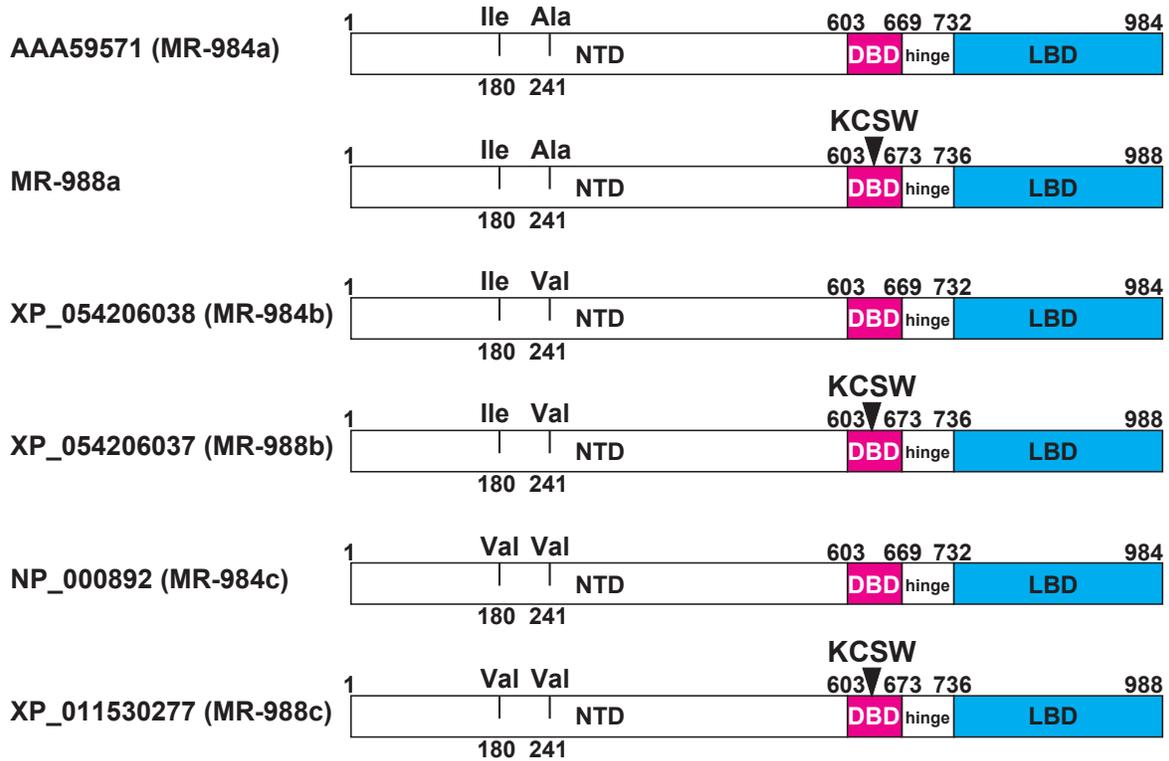

**Figure 2. Human MR Genes in GenBank.** There are three human MR genes with 984 amino acids and two human MR genes with 988 amino acids in GenBank. These MRs contain either isoleucine-180, alanine-241 MR-984a (accession AAA59571), isoleucine-180, valine-241 MR-984b (accession XP_054206038), or valine-180, valine-241 MR-984c (accession NP_000892), in the NTD. GenBank also contains two human MRs with 988 amino acids containing KCSW in the DBD and either isoleucine-180, valine-241 MR-988b (accession XP_054206037), or valine-180, valine-241 MR-988c (accession XP_01530277) in the NTD. A human MR with isoleucine-180 and alanine-241 and a KCSW insert in the NTD was not found in GenBank. MR-988a is our construction of this human MR, which we studied for activation by corticosteroids along with the five human MR genes that are found in GenBank.

Human MR-984a, cloned by Arriza et al [1], has been studied for transcriptional activation by aldosterone and other corticosteroids [1,26,27,29,30]. However, it is not known if the KCSW insert in the DBD of human MR affects its transcriptional activation by corticosteroids, and if so, how? To answer these questions, we compared transcriptional activation by corticosteroids of each human MR-984 to its corresponding human MR-988 transfected into HEK293 cells in the presence of either a TAT3 promoter [42,43] or an MMTV promoter [23,44]. Here we report that compared to human MR-984, human MR-988, has about 50% to 90% increased transcriptional activation by aldosterone, cortisol, and other corticosteroids, when these MRs are transfected into HEK293 cells transfected with a TAT3 promoter. However, in contrast, in HEK293 cells transfected with the MMTV promoter, transcriptional activation by these corticosteroids of each human MR-984 and its corresponding human MR-988 are similar, indicating that the promoter has a role in the regulation of the response to corticosteroids of human MR-988. The physiological responses to corticosteroids in humans with MR genes containing KCSW in the DBD and with differences at codons 180 and 241 in their NTD warrants investigation.

## 2. Results

**Corticosteroid-dependent activation of human MR-984 and human MR-988 containing KCSW.**

We have used human MR-984a (accession AAA59571) [1] for our studies of the human MR [27]. For this project, as described in the Methods Section, we constructed MR-984b (accession XP_054206038), MR-988b (accession XP_054206037), MR-984c (accession NP_000892), MR-988c (accession XP_01530277). We also constructed MR-988a by insertion of KCSW into the DBD of MR-984a. Assays were performed in HEK293 cells transfected with either a TAT3 luciferase reporter (Figure 3, Table 1) or an MMTV luciferase reporter (Figure 4, Table 2).

**MR activation by corticosteroids in the presence of a TAT3 luciferase promoter.**

In Figure 3, we show the concentration dependence of transcriptional activation by corticosteroids of these six human MRs transfected into HEK293 cells with a TAT3 luciferase promoter. Luciferase levels were used to calculate an EC50 and fold-activation for each steroid (Table 1). These results reveal that the three human MR-988 genes with KCSW in the DBD have a substantial increase in transcriptional activation compared to their MR-984 counterparts, which lack KCSW, while there is only a modest change in EC50s of corticosteroids for MR-988 compared to MR-984. As shown in Figure 3 and Table 1, there also were differences in the level of transcriptional activation by corticosteroids of MR-984a, MR-984b and MR-984c (Figure 3A-D), as well as for MR-988a, MR-988b and MR-9848c (Figure 3E-H), which also have different amino acids at codons 180 and 241 in their NTD.

Compared to activation by aldosterone of the three human MRs with 984 amino acids, aldosterone increased fold activation by about 50% for MR-988a, MR-988b and MR-9848c, all of which contain KCSW in the DBD. As shown in Table 1, cortisol, which activates the MR in organs that lack 11β-hydroxysteroid dehydrogenase [7,45–47], increases fold activation by 29% for MR-988a compared to MR984a, by 73% for MR-988b compared to 984b and by 69% for MR-988c compared to MR-984c. Corticosterone increases fold activation by 94% for MR-988a compared to MR984a, by 78% for MR-988b compared to 984b and by 111% for MR-988c compared to MR-984c. 11-Deoxycorticosterone increases fold activation by 94% for MR-988a compared to MR984a, by 78% for MR-988b compared to 984b and by 90% for MR-988c compared to MR-984c.

The different amino acids in human MR at codons 180 and 241 in the NTD also influences corticosteroid activation of transcription. MR-984b and MR-988b, which have Ile-180 and Val-241 in the NTD, have the highest fold-activation, and MR-984c and MR-988c, which have Val-180 and Val-241 in the NTD have the lowest fold-activation. Transcriptional activation by aldosterone, cortisol, corticosterone and 11-deoxycorticosterone of MR-984a, which has Ile-180, Ala-241 in the NTD, is between that of MR-984b and MR-984c. Fold-activation by aldosterone, cortisol and 11-deoxycorticosterone of MR-988a is between that of MR-988b and MR-988c. An exception is corticosterone, which has similar fold-activation of MR-988a and MR-988c. These results indicate that the NTD in human MR has an allosteric influence on transcriptional activation by corticosteroids of human MR.

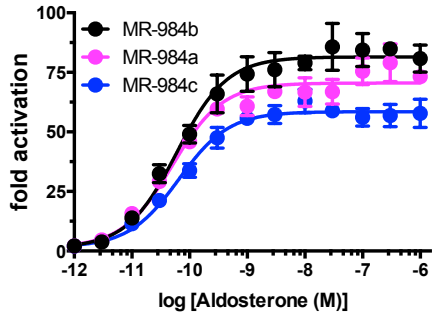

**A: Human MR wild-type with TAT3-luc**

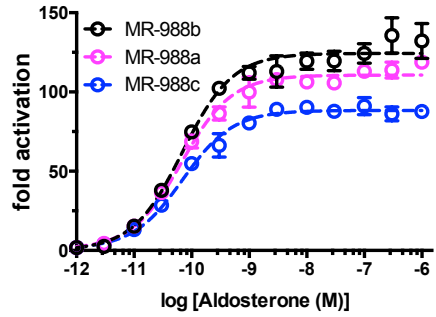

**E: Human MR-KCSW with TAT3-luc**

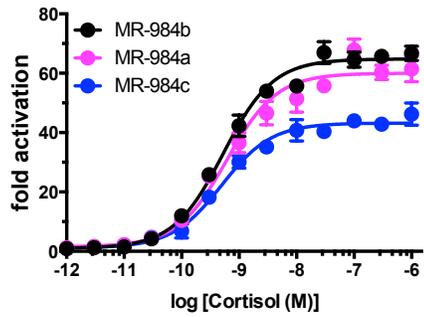

**B: Human MR wild-type with TAT3-luc**

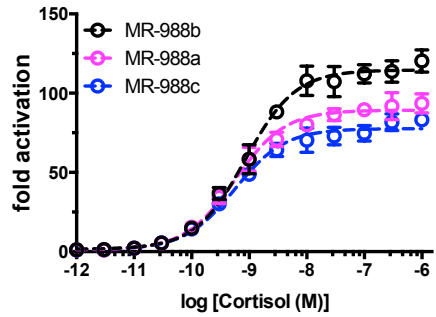

**F: Human MR-KCSW with TAT3-luc**

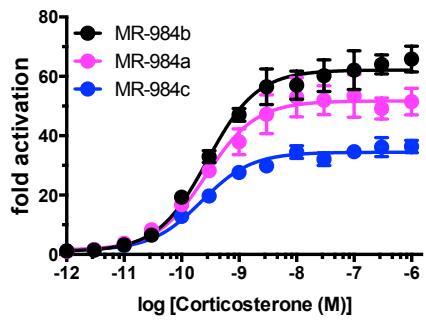

**C: Human MR wild-type with TAT3-luc**

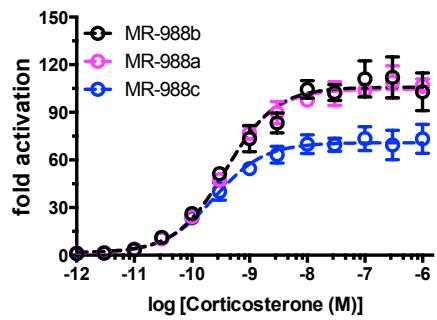

**G: Human MR-KCSW with TAT3-luc**

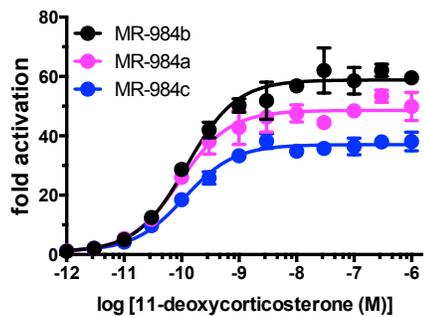

**D: Human MR wild-type with TAT3-luc**

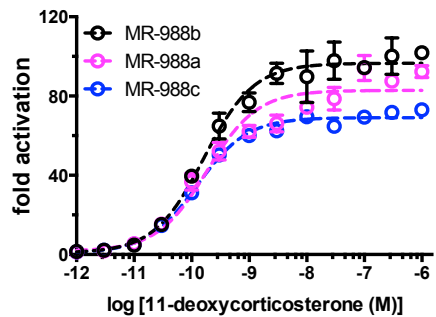

**H: Human MR-KCSW with TAT3-luc**

MR-984a = (Ile-180/Ala-241)
MR-984b = (Ile-180/Val-241)
MR-984c = (Val-180/Val-241)

MR-988a = (Ile-180/Ala-241)
MR-988b = (Ile-180/Val-241)
MR-988c = (Val-180/Val-241)

**Figure 3. Concentration-dependent transcriptional activation by corticosteroids of human MR-984a, b and c and human MR-988a, b and c in HEK-293 cells transfected with a TAT3-luciferase promoter.**

Plasmids for human MR-984a, b, c and human MR-988a, b, c were expressed in HEK293 cells with a TAT3-luciferase promoter [2,42,43]. Cells were treated with increasing concentrations of either aldosterone, cortisol, 11-deoxycorticosterone, corticosterone, or vehicle alone (DMSO). Results are expressed as means ± SEM, n=3. Y-axis indicates fold-activation compared to the activity of vector with vehicle (DMSO) alone as 1. A. Human MR-984 with aldosterone. B. Human MR-984 with cortisol. C. Human MR-984 with corticosterone. D. Human MR-984 with 11-deoxycorticosterone. E. Human MR-988 with aldosterone. F. Human MR-988 with cortisol. G. Human MR-988 with corticosterone. H. Human MR-988 with 11-deoxycorticosterone.

Legend for human MR genes in Figure 3: MR-984a = (Ile-180, Ala-241), MR-984b = (Ile-180, Val-241), MR-984c = (Val-180, Val-241).

**Table 1. Steroid Activation of Human MR in HEK293 Cells with a TAT3 Promoter.**

|  |  | TAT3-luc | | TAT3-luc | | TAT3-luc | |
|---|---|---|---|---|---|---|---|
|  |  | MR-984a | MR-988a | MR-984b | MR-988b | MR-984c | MR-988c |
| Aldosterone | EC50 (nM) | 0.05 | 0.067 | 0.059 | 0.069 | 0.064 | 0.069 |
|  | Fold-Activation (± SEM) * | 75.5 (± 5.61) | 113.1 (± 3.53) | 84.3 (± 6.88) | 124.3 (± 6.23) | 56.0 (± 3.71) | 91.0 (± 5.36) |
|  | Ratio** | 1 | 1.49 | 1 | 1.47 | 1 | 1.62 |
| Cortisol | EC50 (nM) | 0.61 | 0.56 | 0.52 | 0.85 | 0.49 | 0.56 |
|  | Fold-Activation (± SEM) * | 67.8 (± 3.69) | 89.5 (± 3.48) | 64.5 (± 2.55) | 112.1 (± 5.83) | 43.9 (± 1.55) | 74.6 (± 4.85) |
|  | Ratio** | 1 | 1.32 | 1 | 1.73 | 1 | 1.69 |
| Corticosterone | EC50 (nM) | 0.26 | 0.37 | 0.27 | 0.38 | 0.22 | 0.24 |
|  | Fold-Activation (± SEM) * | 53.4 (± 7.21) | 103.9 (± 2.91) | 62.0 (± 6.53) | 111.0 (± 11.35) | 34.6 (± 1.85) | 73.3 (± 7.58) |
|  | Ratio** | 1 | 1.94 | 1 | 1.78 | 1 | 2.11 |
| 11-deoxycorticosterone | EC50 (nM) | 0.092 | 0.18 | 0.12 | 0.16 | 0.11 | 0.12 |
|  | Fold-Activation (± SEM) * | 48.4 (± 1.41) | 94.1 (± 6.28) | 58.5 (± 4.49) | 94.3 (± 1.19) | 36.4 (± 1.64) | 69.3 (± 2.31) |
|  | Ratio** | 1 | 1.94 | 1 | 1.61 | 1 | 1.90 |

*Fold activation values were calculated based on the 100 nM values.
**Ratio values were calculated by dividing the fold-activation value of MR-988 by the value of MR-984. MR-984a/988a = (Ile-180/Ala-241), MR-984b/988b = (Ile-180/Val-241), MR-984c/988c = (Val-180/Val-241). MR-988 contains KCSW motif in DBD.

**MR activation by corticosteroids in the presence of an MMTV luciferase promoter.**

In Figure 4, we show the concentration dependence of transcriptional activation by corticosteroids of six human MRs transfected into HEK293 cells with an MMTV luciferase promoter. Luciferase levels were used to calculate an EC50 and fold-activation for each steroid (Table 2). Each corticosteroid had a similar fold-activation for human MR-984 and its corresponding human MR-988 in assays with an MMTV luciferase reporter, while there is only a modest change in EC50s of corticosteroids for MR-988 compared to MR-984.

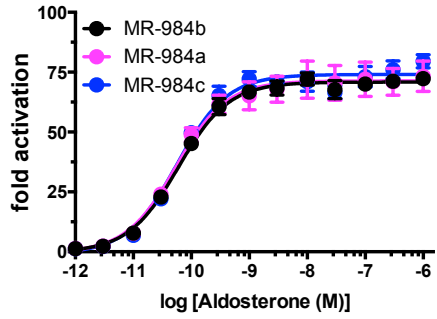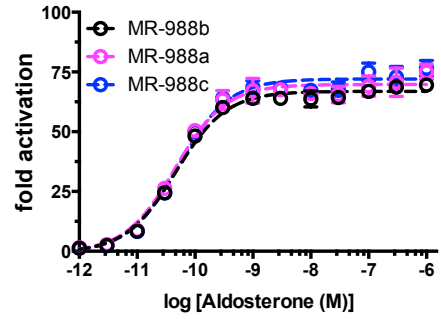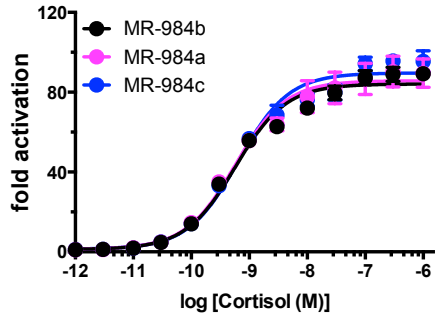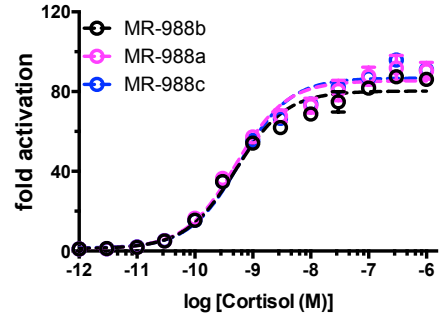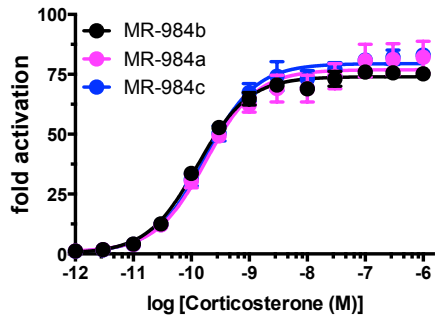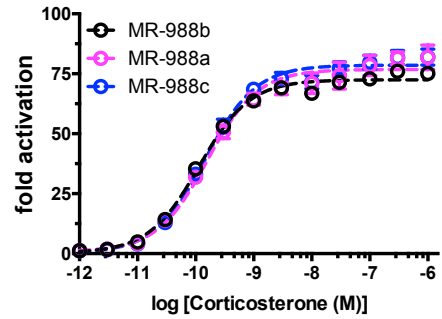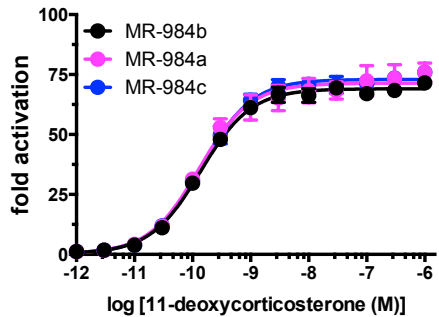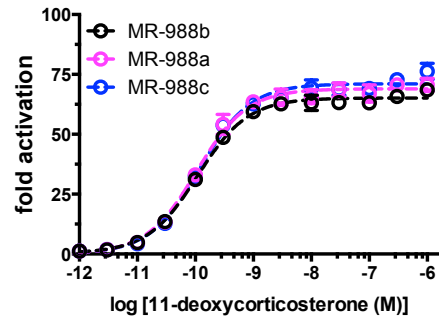

**MR-984a = (Ile-180/Ala-241)**
**MR-984b = (Ile-180/Val-241)**
**MR-984c = (Val-180/Val-241)**

**MR-988a = (Ile-180/Ala-241)**
**MR-988b = (Ile-180/Val-241)**
**MR-988c = (Val-180/Val-241)**

**Figure 4. Concentration-dependent transcriptional activation by corticosteroids of human MR-984a, b and c and human MR-988a, b and c in HEK-293 cells transfected with an MMTV-luciferase promoter.**

Plasmids for human MR-984a, b, c and human MR-988a, b, c were expressed in HEK293 cells with an MMTV-luciferase promoter.  Cells were treated with increasing concentrations of either aldosterone, cortisol, 11-deoxycorticosterone, corticosterone, or vehicle alone (DMSO).  Results are expressed as means ± SEM, n=3.  Y-axis indicates fold-activation compared to the activity of vector with vehicle (DMSO) alone as 1.  A. Human MR-984 with aldosterone.  B. Human MR-984 with cortisol.  C. Human MR-984 with corticosterone.  D. Human MR-984 with 11-deoxycorticosterone.  E. Human MR-988 with aldosterone.  F. Human MR-988 with cortisol.  G. Human MR-988 with corticosterone.  H. Human MR-988 with 11-deoxycorticosterone.

Legend for human MR genes in Figure 4: MR-984a = (Ile-180, Ala-241), MR-984b = (Ile-180, Val-241), MR-984c = (Val-180, Val-241).

**Table 2. Steroid Activation of Human MR in HEK293 Cells with an MMTV Promoter.**

|  |  | MMTV-luc | | MMTV-luc | | MMTV-luc | |
|---|---|---|---|---|---|---|---|
|  |  | MR-984a | MR-988a | MR-984b | MR-988b | MR-984c | MR-988c |
| Aldosterone | EC50 (nM) | 0.054 | 0.045 | 0.06 | 0.046 | 0.057 | 0.052 |
|  | Fold-Activation (± SEM) * | 72.0 (± 7.19) | 69.0 (± 4.27) | 70.0 (± 1.82) | 66.8 (± 0.73) | 73.2 (± 4.31) | 75.0 (± 3.7) |
|  | Ratio** | 1 | 1.04 | 1 | 0.95 | 1 | 1.02 |
| Cortisol | EC50 (nM) | 0.57 | 0.51 | 0.58 | 0.51 | 0.65 | 0.59 |
|  | Fold-Activation (± SEM) * | 86.7 (± 7.77) | 87.7 (± 4.5) | 87.2 (± 3.51) | 81.8 (± 2.68) | 94.2 (± 3.36) | 86.9 (± 3.52) |
|  | Ratio** | 1 | 0.99 | 1 | 0.96 | 1 | 0.92 |
| Corticosterone | EC50 (nM) | 0.18 | 0.16 | 0.13 | 0.12 | 0.17 | 0.15 |
|  | Fold-Activation (± SEM) * | 80.7 (± 6.89) | 78.5 (± 4.23) | 76.0 (± 1.25) | 72.9 (± 0.7) | 81.0 (± 1.79) | 79.8 (± 2.89) |
|  | Ratio** | 1 | 1.03 | 1 | 0.94 | 1 | 0.99 |
| 11-deoxycorticosterone | EC50 (nM) | 0.13 | 0.1 | 0.14 | 0.11 | 0.14 | 0.12 |
|  | Fold-Activation (± SEM) * | 72.4 (± 6.31) | 67.0 (± 3.65) | 67.1 (± 2.21) | 63.1 (± 1.27) | 71.2 (± 2.44) | 69.1 (± 1.17) |
|  | Ratio** | 1 | 1.08 | 1 | 0.94 | 1 | 0.97 |

*Fold activation values were calculated based on the 100 nM values.

**Ratio values were calculated by dividing the fold-activation value of MR-988 by the value of MR-984. MR-984a/988a = (Ile-180/Ala-241), MR-984b/988b = (Ile-180/Val-241), MR-984c/988c = (Val-180/Val-241). MR-988 contains KCSW motif in DBD.

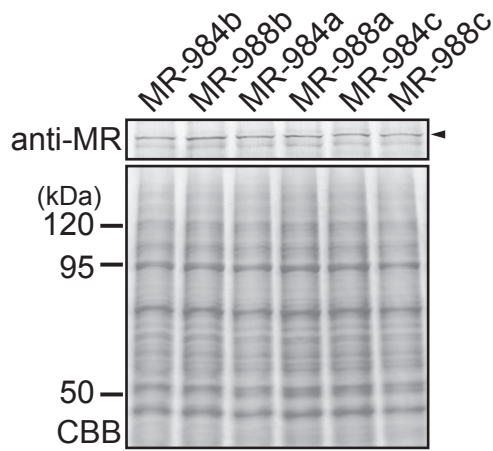

**Figure 5. Western analysis of MR-984a, MR-984b, MR-984c and MR-988a, MR-988b, MR-988c.** HEK293 cell lysates transfected with human MR constructs were treated with sample buffer and applied to a 10% SDS-polyacrylamide gel, and then transferred onto a membrane. The expressed human MR proteins were detected with anti-human MR antibody. Upper panel: Western blot. Lower panel: CBB stain.

Figure 5 shows a Western analysis of the level of each human MR used in the experiments depicted in Figures 3 and 4. The MRs showed no marked difference in protein levels between human MR-984 and the human MR-988. This similarity in protein levels in our assays agrees with the similar fold-activation of the various human MRs in HEK293 cells transfected with an MMTV luciferase promoter shown in Figure 4.

**Discussion.**

The human MR cloned by Arriza et al [1] has 984 amino acids (MR984a), and this is the MR that has been widely studied [7,9–11,13,15,24]. However, our BLAST search of GenBank [41] revealed that MR984a is not the only human MR in GenBank (Figure 2). GenBank also contains two human MRs (MR-988b, MR-988c) with KCSW beginning at codon 634 in the DBD and with different amino acids at codons 180 and 241 in the NTD, than are found in MR-988a, providing total of five distinct human MR sequences. To begin to understand the effects, if any, of the sequence differences among these human MRs, we studied corticosteroid activation of MR-984b, MR-988b, MR-984c and MR-988c for comparison with each other and with MR-984a. We also constructed MR988a for comparison with human MR984a and the other four human MR genes.

We find that corticosteroids have a higher fold-activation for human MR-988 genes compared to their corresponding MR-984 gene in HEK293 cells transfected with a TAT3 promotor, while the half-maximal response (EC50) is similar for corticosteroid activation of MR-984 and MR-988. Transcriptional activation of human MR also depends on the amino acids at codons 180 and 241. However, in HEK293 cells transfected with the MMTV promoter, transcriptional activation by corticosteroids of human MR-988 is similar to activation of human MR-984, which suggests that the promoter has a role in the regulation of the response to corticosteroids of human MR-988.

This increase of MR transcription by aldosterone and other corticosteroids may also occur in other terrestrial vertebrate MRs with KCSW in the DBD (Figure 1). Transcriptional activation of human MR with KCSW is consistent with a homology model of human MR with KCSW in the DBD constructed by Wickert et al. [40]. Their 3D model of human MR-KCSW found no distortion by the four amino acids on adjacent secondary structures of the DBD in human MR. The mechanism(s) for the effects on human MR activity of KCSW and substitutions at codons 180 and 241 are unknown. Recruitment of co-factors and ubiquitination of MRs may be important.

We find that mutations at codons 180 and 241 in the NTD of human MR also affect transcriptional activation by aldosterone and other corticosteroids. This allosteric effect of two mutations in the human NTD was surprising. Compared to chimpanzee, gorilla and orangutan MRs, human MRs are unique in containing an MR with either (Ile-180, Ala-241) or (Val-180, Val-241) in their NTD [41]. That is, chimpanzees, gorillas and orangutans lack an MR with either (Ile-180, Ala-241) or (Val-180, Val-241) in their NTD. We propose that MRs with (Ile-180, Ala-241) or (Val-180, Val-241) in their NTD evolved in humans after their divergence from chimpanzees. In the light of the diverse physiological functions of the MR in humans in regulating transcription in kidney, brain, heart, skin, lungs and other tissues [11,13–22,48], the evolution of human MRs with (Ile-180, Ala-241) or (Val-180, Val-241) in their NTD may have been important in the evolution of humans from chimpanzees [41].

The interaction of the MR with the GR to form heterodimers is important [49–55]. The response to aldosterone and other corticosteroids to heterodimers of human MRs containing KCSW in the DBD, as well as either (Ile-180, Ala-241), (Ile-180, Val-241) or (Val-180, Val-241) in the NTD with human GR needs to be explored.

**Materials & Methods**

**Construction of plasmid vectors**

Full-length mineralocorticoid receptor (MR) of human as registered in Genbank (accession number: XP_054206038) was used as MR-984 to construct MR-984b, MR-984c, MR-988a, MR-988b, MR-988c. The insertion of 4-amino acids (Lys-Cys-Ser-Trp) into the DBD after codon 633 of human MR was performed using KOD-Plus-mutagenesis kit (TOYOBO). The nucleic acid sequences of all constructs were verified by sequencing.

**Chemical reagents**

Cortisol, corticosterone, 11-deoxycorticosterone and aldosterone were purchased from Sigma-Aldrich. For reporter gene assays, all hormones were dissolved in dimethyl-sulfoxide (DMSO); the final DMSO concentration in the culture medium did not exceed 0.1%.

**Transactivation assays and statistical analyses**

Transfection and reporter assays were carried out in HEK293 cells, as described previously [27]. HEK293 cells were seeded in 24-well plates at $5\times10^4$ cells/well in phenol-red free Dulbecco's modified Eagle's medium supplemented with 10% charcoal-stripped fetal calf serum (HyClone). After 24 hours, the cells were transfected with 100 ng of receptor gene, reporter gene containing the *Photinus pyralis* luciferase gene and pRL-tk, as an internal control to normalize for variation in transfection efficiency; pRL-tk contains the *Renilla reniformis* luciferase gene with the herpes simplex virus thymidine kinase promoter. Each assay had a similar number of cells, and assays were done with the same batch of cells in each experiment. All experiments were performed in triplicate. Promoter activity was calculated as firefly (*P. pyralis*)-lucifease activity/sea pansy (*R. reniformis*)-lucifease activity. The values shown are mean ± SEM from three separate experiments, and dose-response data, which were used to calculate the half maximal response (EC50) for each steroid, were analyzed using GraphPad Prism.

**Western blotting**

Human MRs were transfected as described in "Transactivation assays and statistical analysis". Expressed human MR proteins were separated by SDS-PAGE in 10 % gel, blotted onto an Immobilon membrane (Millipore Corp. MA), and probed with anti-human mineralocorticoid antibody (monoclonal mouse IgG, clone #385707, R&D Systems, Inc.).

**Funding:** This work was supported by Grants-in-Aid for Scientific Research from the Ministry of Education, Culture, Sports, Science and Technology of Japan (23K05839) to Y.K., and the Takeda Science Foundation to Y.K.

**Competing interests:** The authors have declared that no competing interests exist.

**Contributions**.

Yoshinao Katsu: Investigation, Conceptualization, Supervision, Formal Analysis, Writing – original draft, Writing – review & editing.

Jiawen Zhan: Data curation, Investigation, Methodology.

Michael E. Baker: Conceptualization; Formal analysis, Supervision, Writing – original draft, Writing – review & editing.